\documentclass[aps,prl,twocolumn,showpacs,superscriptaddress,groupedaddress]{revtex4}
\usepackage{epsfig,latexsym,graphicx,dcolumn,bm}
\usepackage{amsmath}
\usepackage{subfigure}
\usepackage{amssymb}
\usepackage{pstricks, nicefrac}
\usepackage{chemarr, color}

\newcommand{\ben}{\begin{equation}}
\newcommand{\een}{\end{equation}}
\newcommand{\bea}{\begin{array}}
\newcommand{\eea}{\end{array}}
\newcommand{\bef}{\begin{figure}}
\newcommand{\eef}{\end{figure}}

\begin{document}
\title{\large \bf Sticky Surfaces: Sphere-Sphere Adhesion Dynamics} 
\author{ {\bf Sarthok Sircar$^*$, John G.~Younger$^{\dagger}$ \& David M.~Bortz$^{* \ddagger}$}\\{\it $^*$Department of Applied Mathematics, University of Colorado, Boulder, CO 80309-0526} \\{\it $^{\dagger}$Department of Emergency Medicine, University of Michigan, Ann Arbor, MI 48109} \\{\it $^{\ddagger}$BioFrontiers Institute, University of Colorado, Boulder, CO 80309-0596}}
\date{\today}

\begin{abstract}

\noindent We present a multi-scale model to study the attachment of spherical particles with a rigid core, coated with binding ligands and in equilibrium with the surrounding, quiescent fluid medium. This class of fluid-immersed adhesion is widespread in many natural and engineering settings. Our theory highlights how the micro-scale binding kinetics of these ligands, as well as the attractive / repulsive surface potential in an ionic medium effects the eventual macro-scale size distribution of the particle aggregates (flocs). The results suggest that the presence of elastic ligands on the particle surface allow large floc aggregates by inducing efficient inter-floc collisions (i.e., a large, non-zero collision factor). Strong electrolytic composition of the surrounding fluid favors large floc formation as well. 
\end{abstract}

\pacs{}
\maketitle


%

\noindent {\it Introduction} -- The formation of aggregates, induced by the adhesion of two spherical particles or nearby surfaces is important in many scientific and industrial processes. Interfacial attachment leading to larger floc aggregates via the latching of binders on surfaces in close proximity is widespread. Examples include binding of bacterial clusters to medical implants and host cell surfaces \cite{Zhu2000}, cancer cell metastasis \cite{Lei1999}, and the coalescence of medical gels with nano-particles for targeted drug delivery \cite{Moore2006}. Moreover, coagulation and flocculation (the chemical and the physical aspects of adhesion) are also important in pulp and paper-making industries as well as wastewater treatment plants \cite{Somasundaran}.  In particular, in the wastewater literature, the microscale model for adhesion between all sizes of aggregates is between two rigid, spherical bacteria, regardless of the number and shape of the aggregate and its consitutive bacteria \cite{Maximova2006,Vogelaar2005}. The microscale model is then upscaled to a population model for the design and management of wastewater treatment plants \cite{Mohapatra2010}.  This model under-predicts observed levels of aggregation \cite{Vogelaar2005} and thus is one phenomena (amongst the many described in this section) which would benefit from a more first-principles approach to modeling ligand-mediated rigid sphere-sphere adhesion.

Past investigations in the micro-scale modeling of fluid-borne surface adhesion have addressed many theoretical challenges. These include the ligand-receptor binding kinetics \cite{Dembo1988, Bell1978}, particle surface deformation \cite{Reboux2008, Evans1985} and flow past the surrounding surfaces \cite{GoldmanCoxBrenner1967CES, King2001}. Consequently, many detailed kinetic models have successfully described the adhesion-fragmentation processes from the microscopic perspective. Schwarz \cite{Korn2006} and more recently Mahadevan \cite{Mani2012} studied the cellular adhesion between the ligand coated wall and a sphere moving in a shear flow. A similar model by Seifert described the membrane adhesion via Langevin simulations \cite{Bihr2012}. On the contrary, the macro-scale phase-field models describe the geometry of the floc aggregates as a continuum mass of EPS (extracellular polymeric substance) and predict the stability of the anisotropic structures in a flowing medium. For example, Keener found that in one dimensional space, the polymer-solvent interface is unstable in a flowing medium and phase separate in a finite time interval \cite{Cogan2004}. In two dimensional space, these instabilities spawn several peculiarities including rippling, streaming, merging and detachment \cite{Zhang2008}.

However, efforts to couple the microscopic ligand evolution kinetics of charged surfaces with the macroscopic population model of particle aggregation dynamics are limited. Sciortino made a recent effort in this direction, but those numerical studies were done for a very limited case of aggregation (e.g., irreversible aggregation) and with chemically inert particles \cite{Corezzi2012}. Further, in many population models, the rate of adhesion is typically described as a simple product between aggregate sizes, e.g., $\kappa' xy$ \cite{Somasundaran}.  The coefficient $\kappa'$ is an \emph{adhesion efficacy} which is fit to experimental data.  This rate is an extremely imprecise characterization of the multitude of factors influencing adhesion.

This article represents an initial effort toward resolving this imprecision by carefully combining the micro-scale description of the overall aggregation rate with the macro-scale floc-size distribution. We explore how this adhesion ({\it collision} as termed in the colloid science literature) mechanism for rigid, micron-size, spherical flocs is governed by various geometric and fluid parameters as well as how the surface forces and binding kinetics of the ligands impact the eventual size of these flocs. We consider the sphere-sphere interactions in a quiescent (or no-flow) fluid conditions. The microscopic description of ligand-mediated surface adhesion is an important case from an experimental point of view, e.g., consider the experiments by Sokurenko et al.~which studies the fimbriae mediated catch bond interactions of {\it E.~coli} in stagnant conditions \cite{Sokurenko1997, Sokurenko1998}. While {\it E.~coli} are typically rod-shaped, these bacteria can become coccoid under common environmental response / non-lethal mutations \cite{Cooper1997}. Moreover, our approximation of the spheres as rigid is relevant to both Gram-negative and Gram-positive bacteria which can have cell walls with longitudinal Young's moduli of 200MPa \cite{Tuson2012}. Another example is the experimental studies of the P-selectin/PSGL-1 catch bond interactions of leukocytes (a roughly spherical particle) with and without fluid flow \cite{Marshall2003, Thomas2008}. The rigid microspheres in these case studies had much shorter bonds (no microvilli) and higher spring stiffness.

\noindent {\it Model} -- The present study is geared towards tracking the size distribution of the floc aggregates in equilibrium with the surrounding stagnant fluid \cite{Byrneetal2011PRE}. The spherical particles within the flocs adhere through well-defined disc-like patches covered with binding ligands. Following the general outline given in \cite{BortzEtal2008bmb}, we define $b(t, x) \triangle x$, as the number of aggregates having volumes between $x$ \& $x + \triangle x$ in time $t$. In the volumes between $x_1$ and $x_2$, the total number of flocs $B_0$ is given by 
\ben
B_0(t, x_1, x_2) = \int^{x_2}_{x_1} b(t, x) d x
\een
for [$x_1, x_2$] $\subset$ [$\underline{x}, \overline{x}$], where $\underline{x}$ and $\overline{x}$ are the minimum and maximum aggregate volume sizes, respectively. A finite nutrient supply and the duration of the experiment allow us to assume that $\overline{x}$ is finite. Further, the minimal size $\underline{x}$ is the volume of one particle.
%
%
The conservation of the aggregate number density, or the governing equation for $b$ is \cite{BortzEtal2008bmb}
\ben
b_t = A_{\text{in}}(x, b) - A_{\text{out}}(x, b) \label{eq:Smol_final}
\een
where $A_{\text{in}}$ is the rate with which flocs of size in [$x,x + \triangle x$] are created and $A_{\text{out}}$ is the rate a floc of size in [$x,x + \triangle x$] joins with another floc, to form a floc of volume greater than $x + \triangle x$. These rates are given by
\begin{subequations}
\begin{align} 
A_{\text{in}} (x, b) & = \frac{1}{2} \int^{x-\underline{x}}_{\underline{x}} K_A(y, x-y) b(t,y) b(t,x-y) dy, \nonumber  \\
& \hspace{4cm} x \in [2\underline{x}, \overline{x}] \label{eq:Ain} \\
A_{\text{out}} (x, b) &= b(t,y) \int^{x-\underline{x}}_{\underline{x}} K_A(x,y)  b(t,y) dy, \nonumber  \\ 
& \hspace{3.3cm} x \in [\underline{x}, x-\underline{x}] \label{eq:Aout}
\end{align} 
\end{subequations}
$K_A$ is the aggregation kernel, describing the rate with which flocs of volume $x$ and $y$ combine to form a floc of volume $x + y$. The next two sections will focus on modeling this kernel based on (a) the surface binding kinetics and (b) surface potential, of two coalescing, charged spherical floc-surfaces. \vskip 10pt

\begin{figure}[htbp]
\centering
\includegraphics[scale=0.24]{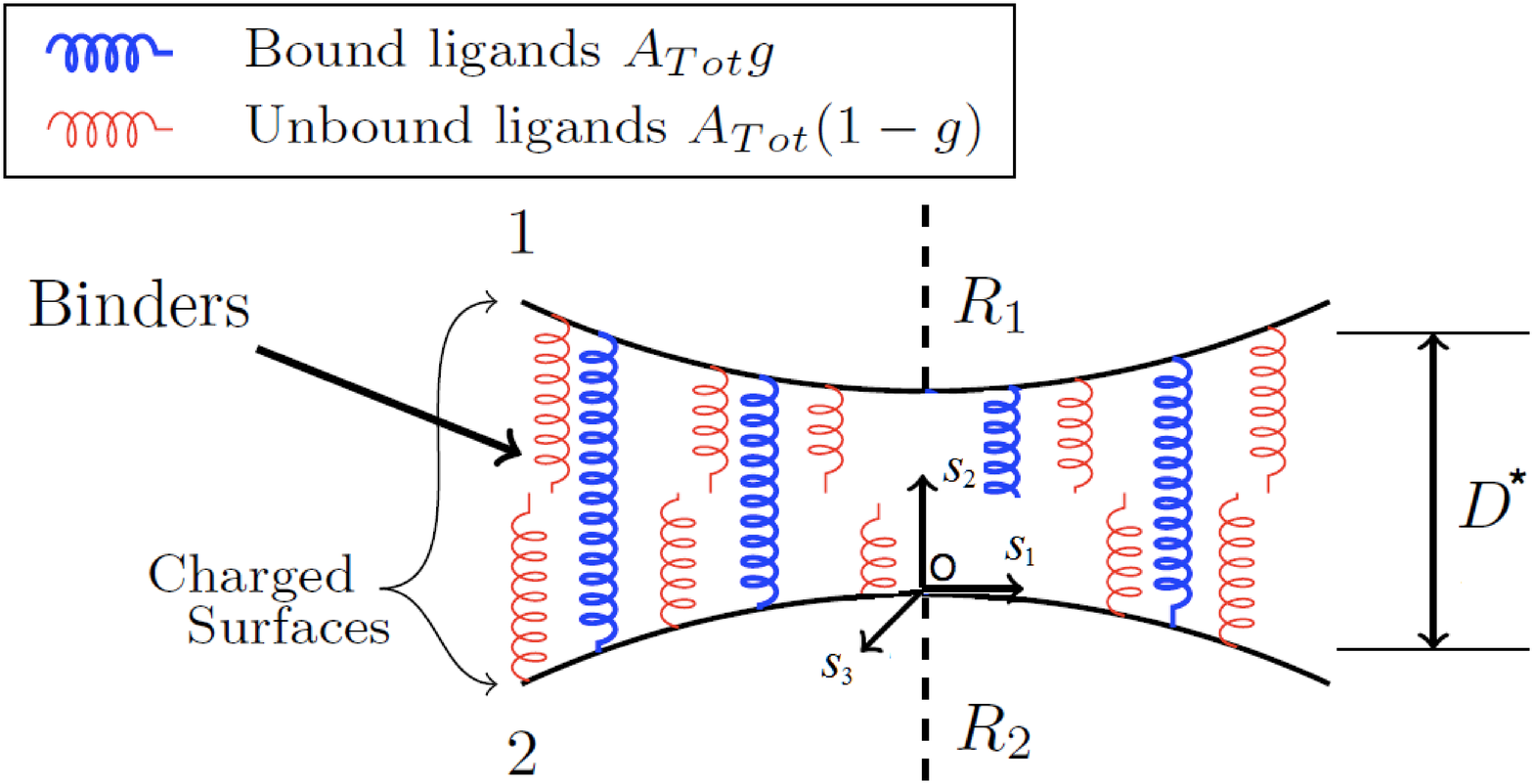}
\vskip -10pt
\caption{Two coalescing spherical flocs coated with binding ligands.}\label{fig:ad_illus}
\end{figure}

\noindent {\it (a) Surface binding kinetics} -- Fig.\ref{fig:ad_illus} illustrates a model of interfacial attachment between two spheres of radius R$_1$ and R$_2$, immersed in a stagnant fluid medium \cite{Dembo1988}. The center, O, of the local spatial frame is located at the point of minimum separation and on the surface of sphere 2. The surface of the spheres bind onto each other due to the presence of adherent elastic binders (which are polymer strands with sticky heads) attached on the floc surfaces. The floc core does not deform. The binders are idealized as linear Hookean springs with stiffness $\kappa_0$ and mean rest length ${\it l}_0$. 

For a given spatial point ${\bf s} = (s_1, s_2, s_3)$ with respect to the center, O, of the local frame, let $D({\bf s})$ be the separation distance between the two spheres, $A_{\text{Tot}}$ and $g$ be the total number of binding ligands and the density of bound ligands on the adhesion surface, respectively. For notational simplicity, we denote $D({\bf s}) = D^*$. This separation distance with respect to the local frame is given by, $D^* = \frac{s^2_1 + s^2_3}{2}(\frac{1}{R_1} + \frac{1}{R_2}) + D$, where D is the minimum separation distance (see Fig.\ref{fig:ad_illus}). Next, we define A$_{\text{Tot}} g(t) d A$ as the number of bonds in the transverse direction that are attached between the surfaces $d A$ at time $t$. In floc literature, the function $g$ is synonymous to the collision factor as well. Hence the total number of bonds formed in the transverse direction is $\int_{A_c} A_{\text{Tot}} g(t) d A$, $A_c$ being the area of adhesion \cite{Dembo1988}. The bond attachment / detachment rates, are

\vskip -20pt
\begin{align}
K_{\text{on}} (D) &= K^*_{\text{on}} \exp \big[ \frac{- \kappa_s(D-{\it l}_0)^2 + W(D) }{ 2{\it k}_B T} \big ], \nonumber \\
K_{\text{off}} (D) &= K^*_{\text{off}} \exp \big[ \frac{(\kappa_0 - \kappa_s)(D-{\it l}_0)^2 + W(D)}{2{\it k}_B T} \big], \label{eqn:reaction_rates}
\end{align}
respectively, where {\it k}$_B$ is the Boltzmann constant, T is the temperature, $\kappa_s$ is the spring constant of the transition state used to distinguish catch ($\kappa < \kappa_s$) from slip ($\kappa > \kappa_s$) bonds \cite{Dembo1988}, $W(D)$ is the total surface potential (described in the next section), $K^*_{\text{on, eq}}, K^*_{\text{off, eq}}$ are the equilibrium binding affinities.  Neglecting the competition for binding sites (i.e. in the limit of small binding affinity and abundant binding receptors on the sphere surface or $\nicefrac{A_{\text{Tot}} K_{\text{on, eq}} }{ K_{\text{off, eq}} } \ll$ 1), the spatially independent bond ligand density evolves via the following differential equation \cite{Dembo1988}: 
\ben
\frac{d g}{d t} = A_{\text{Tot}} K_{\text{on}} - K_{\text{off}} g.
\label{eq:coll_fac}
\een
%

\begin{figure}[htbp]
\centering
\subfigure[]{\includegraphics[scale=0.4]{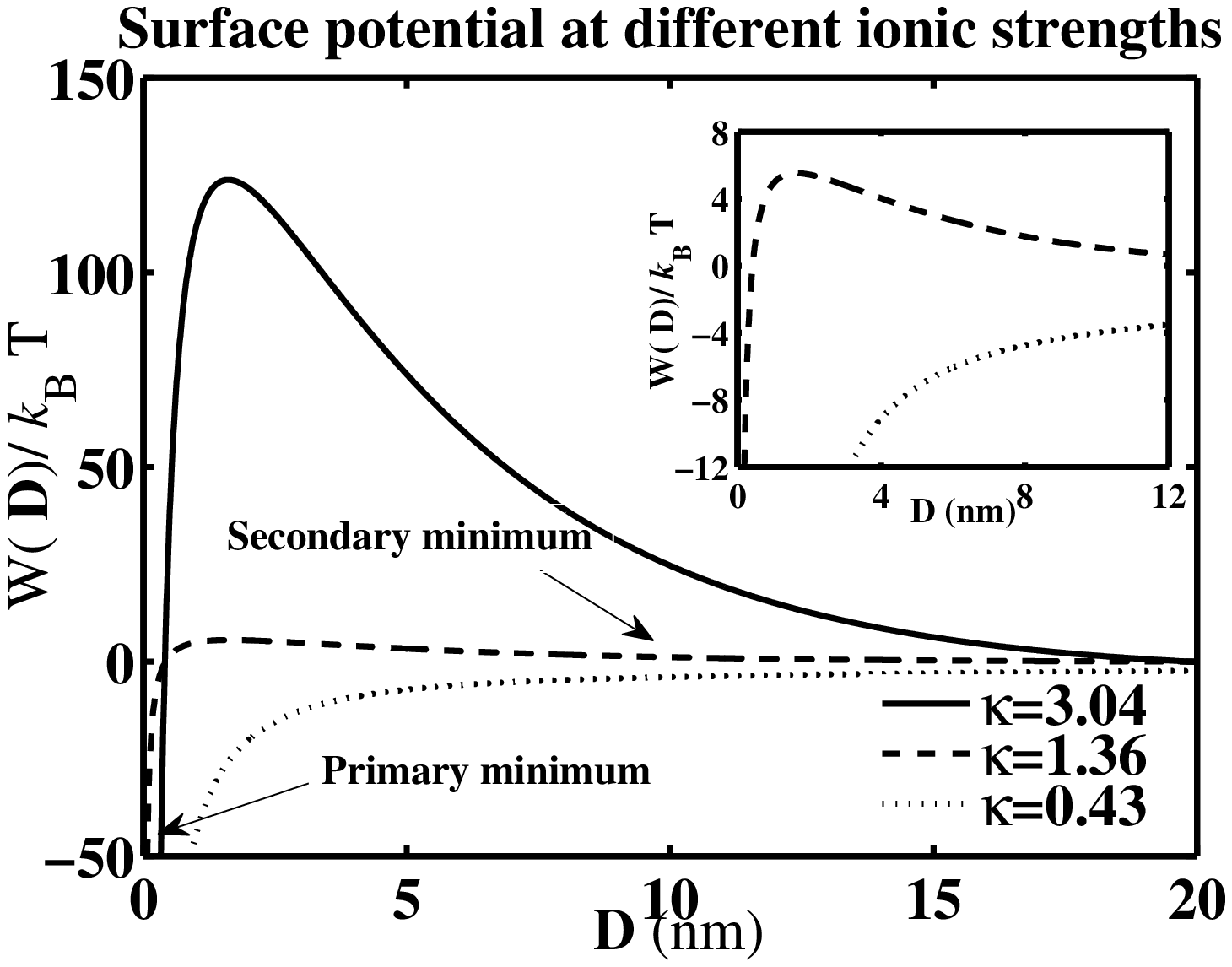}}
\subfigure[]{\includegraphics[scale=0.4]{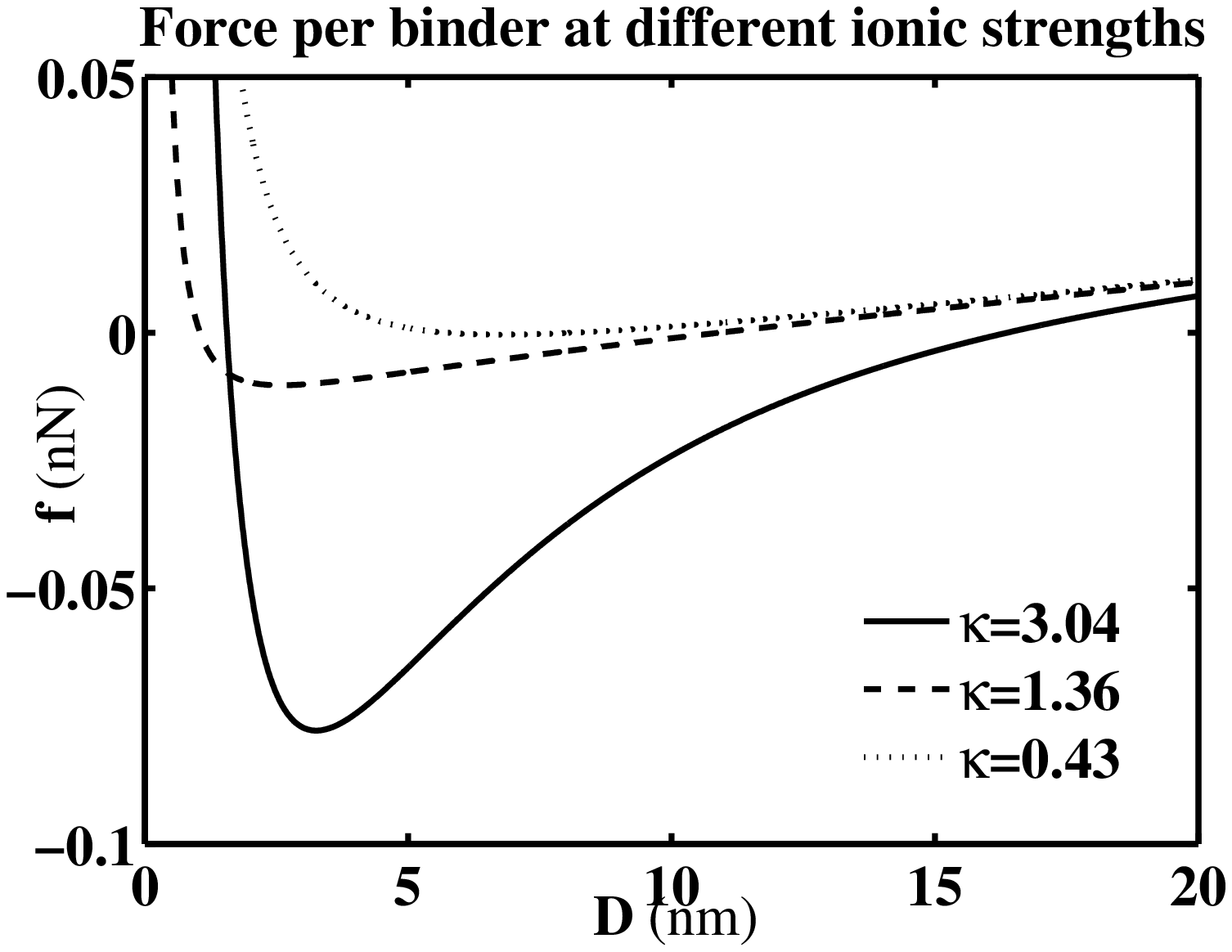}}
\vskip -10pt
\caption{(a) Total surface potential, $W(D)$ versus the separation distance $D$, for two rigid, spherical flocs of radii $R_1=0.25 \mu$m  and $R_2=0.5 \mu$m respectively, and (b) surface force per binder, {\bf f}, (Eq.(\ref{eq:fs})) versus $D$; at different ionic concentration of a 1:1 electrolyte. Regions of attraction: {\bf f} $>$ 0, region of repulsion: {\bf f} $<$ 0.}\label{fig:potential}
\end{figure}

\noindent {\it (b) Surface potential} -- We describe these interactions on the rigid surface of charged spherical flocs, through the DLVO approach, i.e., the Coulombic and Van der Waals interaction. %
%
%
%
%
For two charged spheres, with radii R$_1$, R$_2$, the repulsive Coulombic forces in the gap, D, is given by 
\ben
W_{\text{C}}(D) = 2\pi \epsilon_0 \epsilon \psi_1 \psi_2 \Big(\frac{2R_1 R_2}{R_1 + R_2}\Big) e^{-\kappa D} \label{eq:CI}
\een
where $\kappa$ is the Debye length, $\epsilon, \epsilon_0$ are the dielectric constant of vaccum and the medium, respectively, $\psi_1, \psi_2$ are the {\it zeta potentials} of the respective spheres. The attractive Van der Waal forces for spherical flocs in the regime of close contact ($D \ll R_1, R_2$), is 
%
\ben
W_{\text{VW}}(D) = -\frac{A}{6D} \frac{R_1 R_2}{R_1 + R_2} \label{eq:VWI}
\een  
where A is the Hamaker constant, measuring the Van der Waal `two-body' pair-interaction for macroscopic objects.  
%
%
The net surface potential is $W(D) = W_{\text{C}}(D) + W_{\text{VW}}(D)$, which is pair-wise attractive over very short and very long distances, and pair-wise repulsive over intermediate distances (Fig.\ref{fig:potential}).  \vskip 10pt

Under the influence of the binding kinetics and the surface charges, the instantaneous force that these two colliding charged surfaces exert on each other (and acting normal to the surface) becomes:
\ben
{\it \bf f}(D) = \kappa_0 (D - {\it l}_0) + \nabla_{{\bf D}} \cdot W(D)
\label{eq:fs} \een
where the first term represents the stretching force from the binders due to Hooke's law and the second term represents the forces due to the surface potential. The direction of this force is along the direction vector from the spherical floc of radius R$_2$ towards the floc of radius R$_1$ (Fig.\ref{fig:ad_illus}). The total force arising from all such bonds formed, is given by
\vskip -10pt
\ben
{\bf F}(D,t) = A_{\text{Tot}} \int_{A_c} g(t) {\it \bf f}(D) dA 
\label{eq:Fs} \een
The adhesion area is given by A$_c = \pi$ R$_c^2$. The adhesion radius, $R_c$, is found using an identical scaling law argument of the `settling phase of the particles' (supp. mat.  \cite{Mani2012}), and is given by 
\ben
R_c = \Big(\frac{2{\it k}_B T}{\kappa_0} \Big)^{1/4} {\it l}_0 (R^{-1/2}_1 + R^{-1/2}_2).
\een
%
To summarize, we have presented a model describing the surface adhesion of round flocs, with the following salient features: \vskip 2pt

\noindent $\bullet$ Each floc constitutes a rigid spherical core onto which linear, hookean, spring-like binding ligands are attached and the surface of the coalescing flocs are linked through these ligands. The ligand kinetics is modeled using a differential equation mediated by the bond formation/breakage rates. \vskip 2pt

\noindent $\bullet$ The rigid core of the floc is charged and suspended in an ionic medium. The charge effects are modeled via the repulsive Coulombic interactions and the attractive Van der Waal interactions. \vskip 10pt

%
%
%
\noindent {\it Numerical results} -- Some limitations are imposed in our current approach. In stationary fluid conditions, the hydrodynamic interactions and the spatial inhomogeneities are absent, and this allows the binders to attach/detach normal to the adhering surface (thereby ignoring the tangential displacement of the spheres). Further, the binder kinetics is assumed to be independent of the salt concentration (i.e. the spring stiffness, $\kappa_0$ is independent of the charge-screening length, $\kappa$. This implies that we are neglecting the electro-viscous stresses \cite{Tabatabaei2010}). Compared with the time scale of floc aggregation (or the time scale on which the aggregate number density changes), the attachment / detachment rates of the flocs is sufficiently rapid so that the non-equilibrium binding kinetics can be ignored (i.e., $\frac{d g}{d t}$ = 0 in Eqn.(\ref{eq:coll_fac})). This last assumption may not be realistic in some cases but some groups have shown that the results are, otherwise, qualitatively similar \cite{Reboux2008}. The bond-ligand density (or the collision factor), $g$, (Eqn.~(\ref{eqn:reaction_rates}, \ref{eq:coll_fac})) then evolves according to 
%
\ben
g = A_{\text{Tot}} \frac{K_{\text{on}}}{K_{\text{off}}} = \frac{A_{\text{Tot}} K^*_{\text{on}}}{K^*_{\text{off}}} e^{-\kappa_0\frac{(D-{\it l}_0)^2}{2{\it k}_B T}}, \label{eq:g_symm}
\een
and the total adhesion force between two flocs (Eqn.(\ref{eq:Fs})) reduces to 
\ben
{\bf F}(D) = A_{\text{Tot}}g {\bf f} [ \pi R^2_c ] \label{eq:TotalForce}
\een
Finally, in a Stokes flow, the aggregation rate, K$_A$ (defined in the aggregate number density Eqn.(\ref{eq:Smol_final})), is 
\ben
K_A = \nicefrac{\gamma_A A_c {\bf F} }{ \zeta}
\label{eq:KA} \een
$\zeta$ is the drag coefficient, $\gamma_A$ is the aggregation contact efficiency parameter. Eqns. (\ref{eq:Smol_final}, \ref{eq:Ain}, \ref{eq:Aout}, \ref{eqn:reaction_rates}, \ref{eq:CI}, \ref{eq:VWI}, \ref{eq:fs}, \ref{eq:g_symm}, \ref{eq:TotalForce}, \ref{eq:KA}) along with initial conditions, $b(0, x)=b_0(x)$, constitutes the entire system which calculates the size distribution of round floc aggregates.
%

\noindent {\it (a) Surface potential and bond density calculations} -- The pair-wise surface potential, $W(D)$, is valid over relatively short distances ($D \ll R_1, R_2$). The salt dissolved in the fluid is assumed to be a 1-1 electrolyte at different concentrations, zeta potentials and Debye lengths. These values are used from Camesano's experiments involving adhesion of rigid spherical bacterial surface with silicon nitride AFM tip \cite{Abu-Lail2003}, and are listed in Table \ref{tab:DLVO}. The calculation of the Debye length from different electrolyte concentration is given in \cite{Israelachvili2011} (Chap-14). The dielectric constant in vaccum is $\epsilon_0 = 8.854 \times  10^{-12}$, while the permittivity of water at T=25$^o$C is $\epsilon=78.5$. The Hamaker constant measuring the macroscopic Van-der Waal sphere-sphere interaction is fixed at 2.44 {\it k}$_B$T \cite{Gregory2006}.
\begin{table}[htbp]
\centering
\begin{tabular}{|c|c|c|c|c|}
\hline
[salt] (M) & $\psi_1(mV) $ & $\psi_2$ (mV) & $\kappa$\\
\hline
$ 0.01 $ & -16 & -31.7 & 3.04 \\
\hline
0.05 & -14 & -9.2 & 1.36 \\
\hline
0.5 & -10 & -3 & 0.43 \\
\hline
\end{tabular}
\caption{Parameters corresponding to the Coulombic interactions, Eqn.(\ref{eq:CI}).}\label{tab:DLVO}
\end{table}

\begin{figure}[htbp]
\centering
\subfigure[]{\includegraphics[scale=0.39]{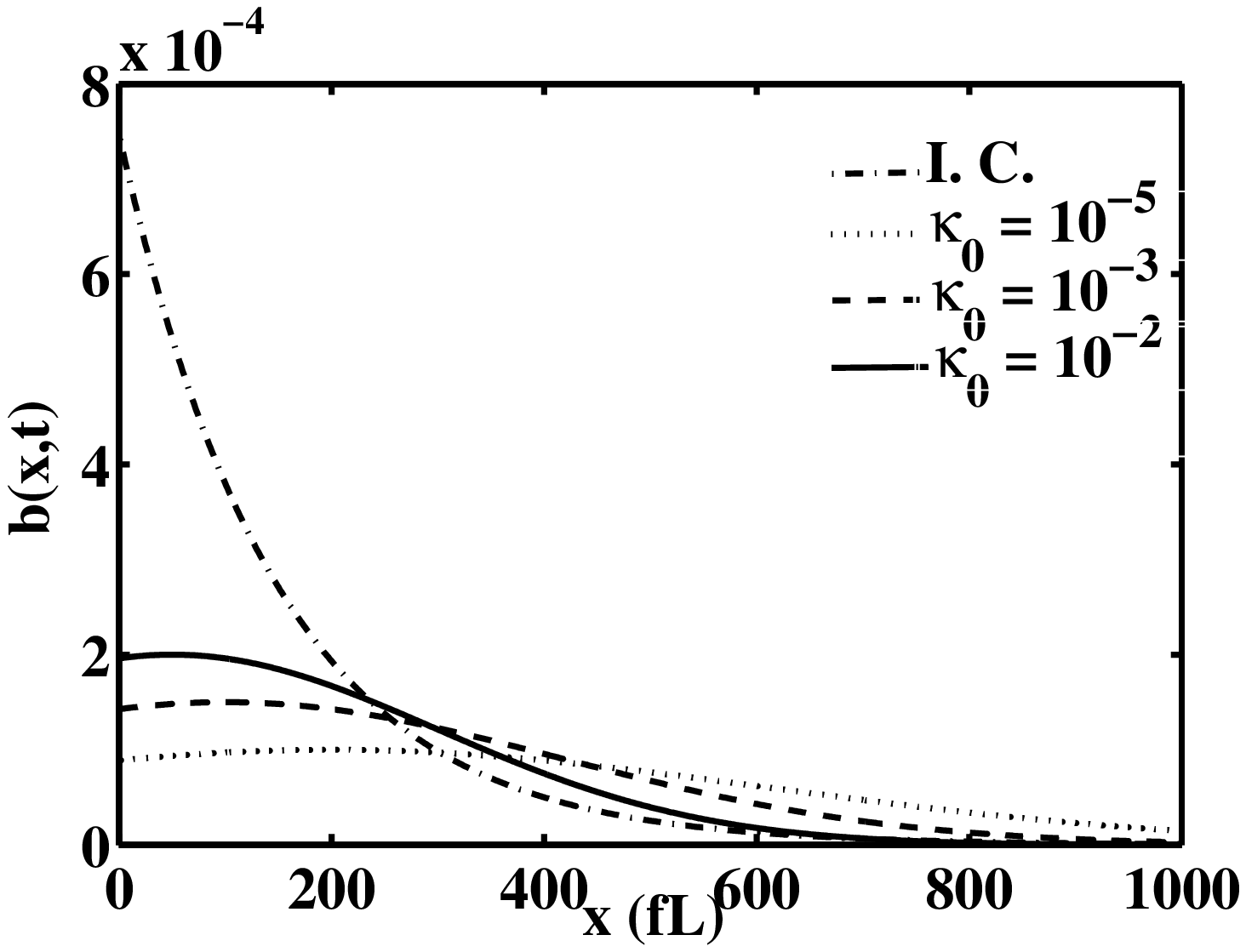}} \vskip -2pt
\subfigure[]{\includegraphics[scale=0.39]{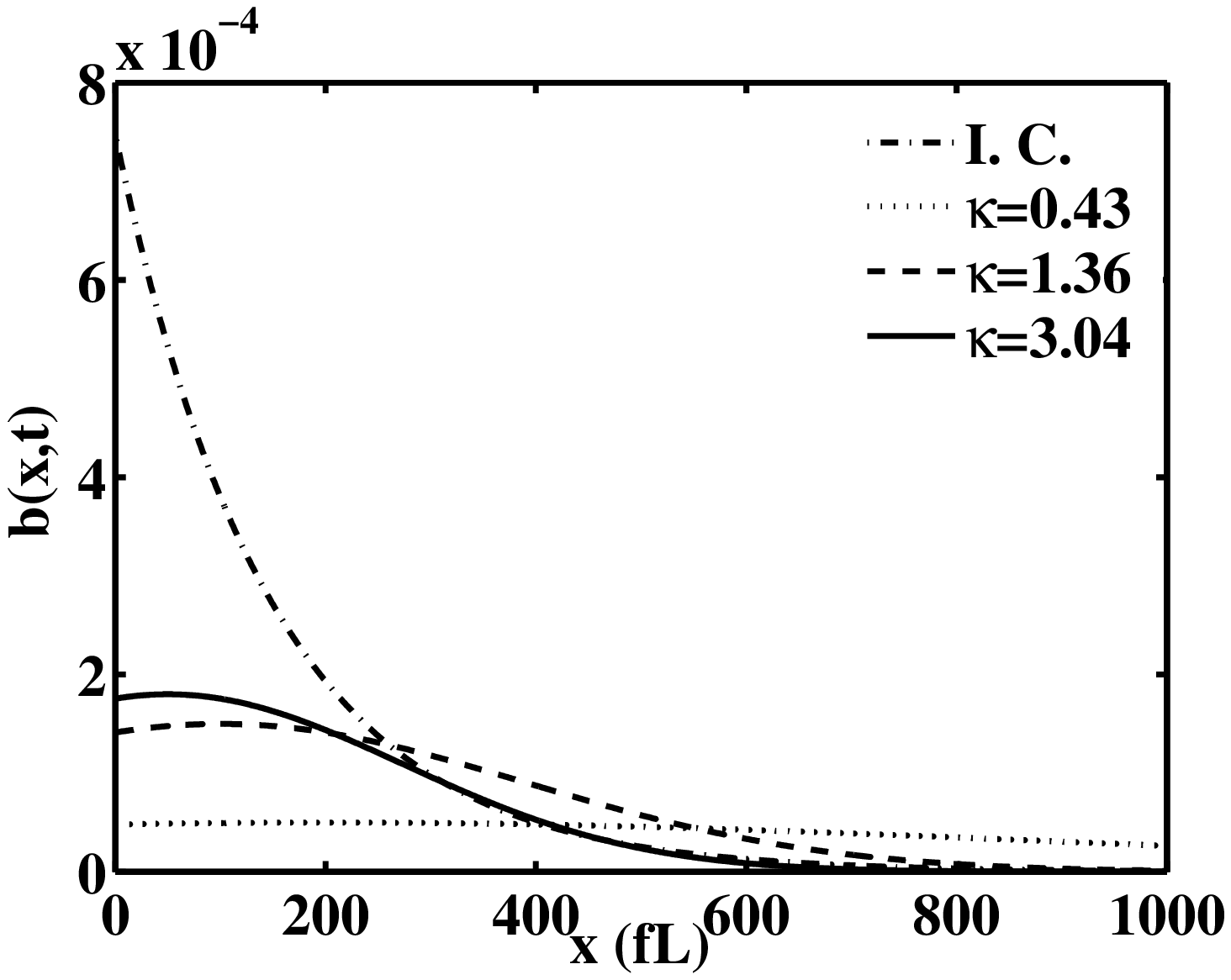}}
\vskip -10pt
\caption{Floc number density distribution versus floc-volume at time T=100 min for (a) different binder stiffness and screening length, $\kappa=1.5$, and (c) different screening lengths and $\kappa_0 = 10^{-3}$ Nm$^{-1}$. The dash-dot curve in these figures is the initial conditions (b$_0$(x)).}\label{fig:b} 
\end{figure}

A weak electrolytic solution (e.g., $\kappa=3.04$ curve, Fig. \ref{fig:potential}a) has a large potential energy barrier at short separation distances, since a weak salt solution results in diffuse screening length surrounding the charged surfaces which hinders adhesion (see the floc population studies in Fig. \ref{fig:b}a). Conversely, for sufficiently concentrated solution (e.g., $\kappa=0.43$ curve, Fig. \ref{fig:potential}a), the energy barrier is reduced and aggregation is favored. The {\it primary minima} (shown in Fig. \ref{fig:potential}a) is unphysical, since at very short separation distances the non-DLVO repulsive steric interaction is dominant and that prevents the surface of the particles from coming into true contact. The regions of attraction/repulsion of this potential is inferred from surface force per binder, {\bf f} (Fig. \ref{fig:potential}b). For sufficiently concentrated salt solution these forces are attractive ({\bf f} $> 0$ for all $D$, $\kappa=0.43$ curve, Fig.\ref{fig:potential}b) and hence, adhesion is always favored. Otherwise at lower salt concentrations, the general feature is that at intermediate distances (2nm $<$ D $<$ 15 nm), the short-range repulsive Coulombic forces are dominant while at longer distances (D $>$ 15 nm), the adhesive forces are dictated by the attractive spring force of the stretched binders. We choose to conduct our numerical simulations for calculating floc-size distribution at a minimum separation distance $D = 11$nm, a point far away from the {\it primary minima} where the adhesive forces are attractive. 
%

%
%

\begin{figure}[htbp]
\centering
\includegraphics[scale=0.39]{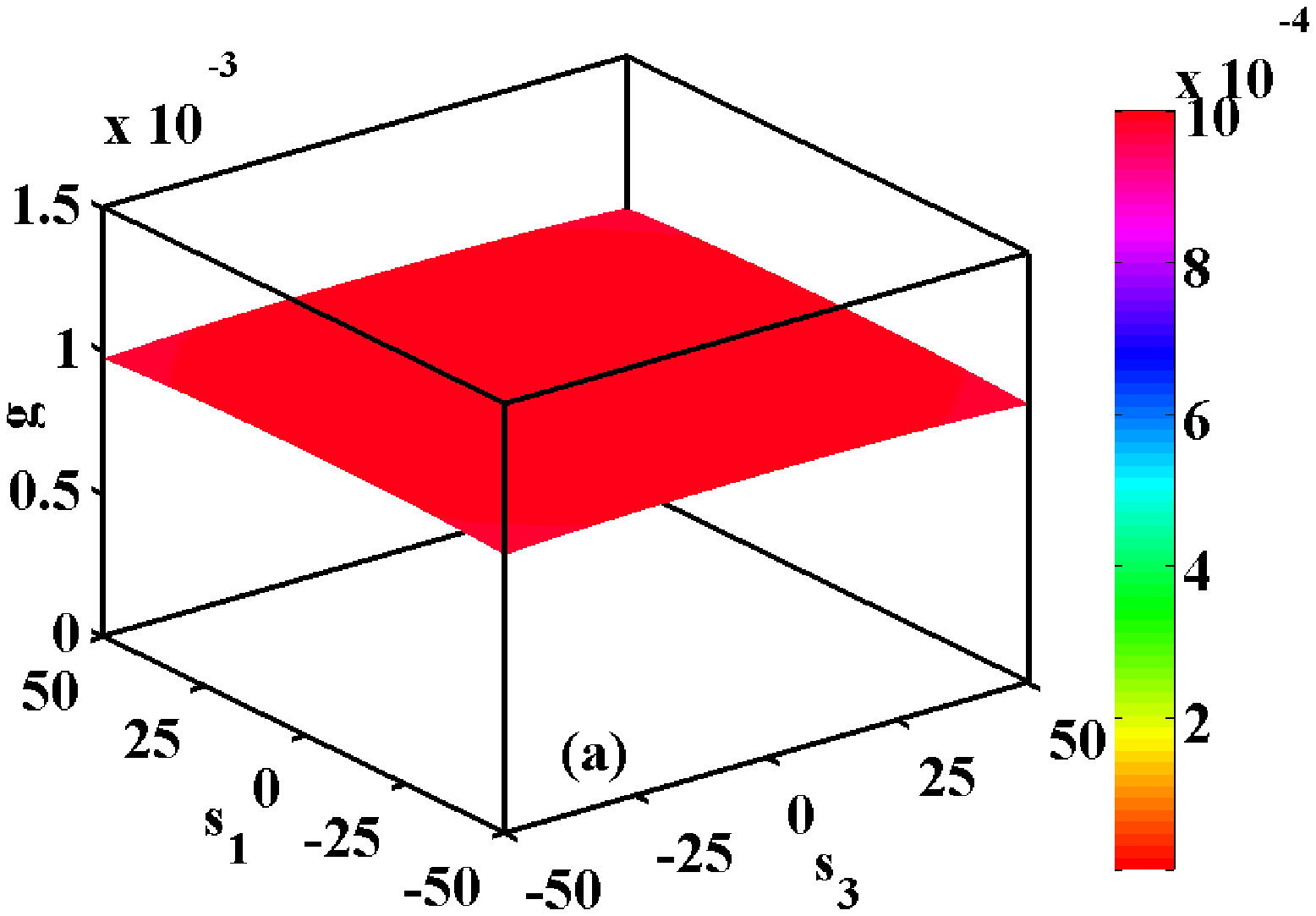}
\includegraphics[scale=0.39]{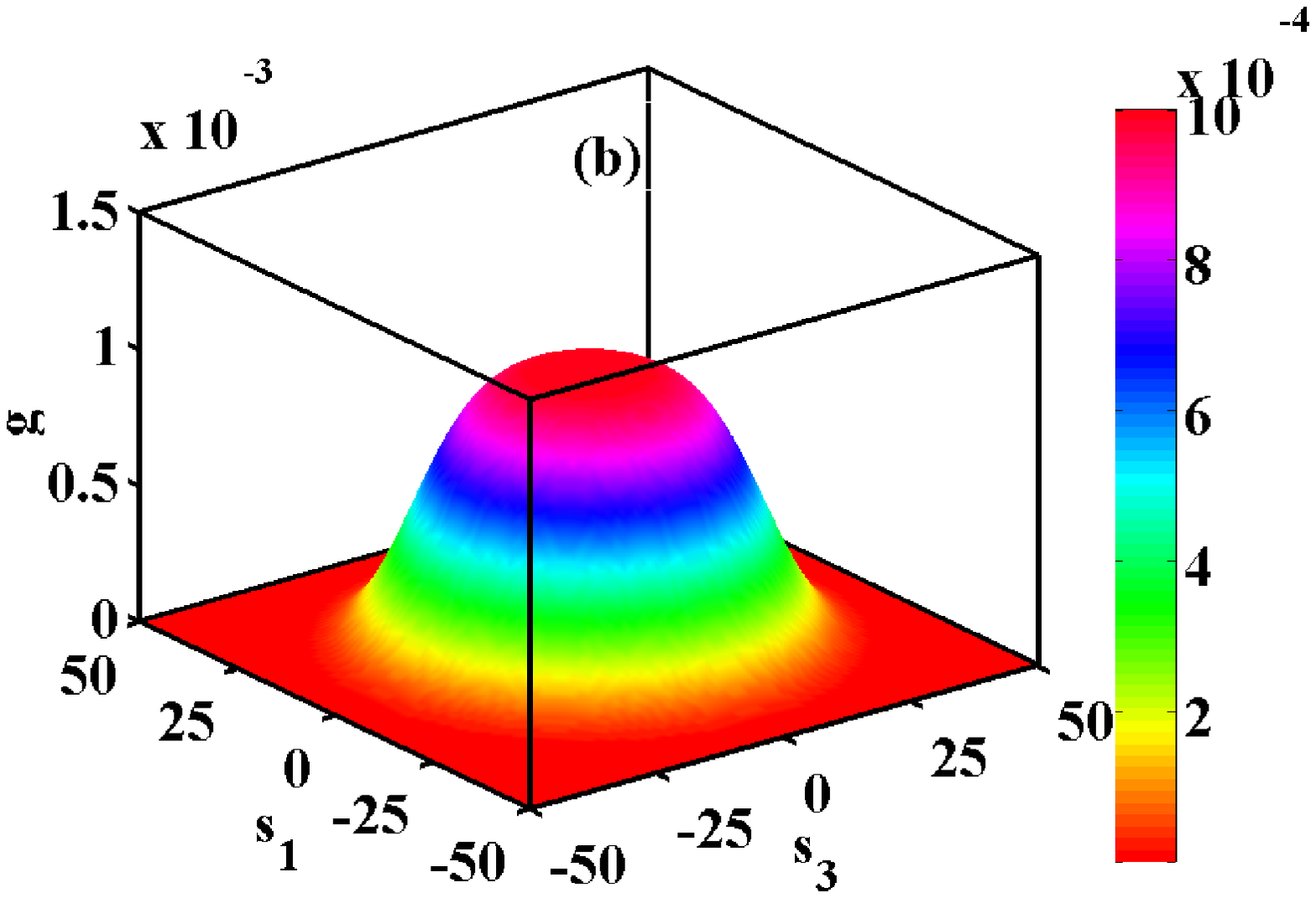}
\vskip -10pt
\caption{Collision factor $g$ vs spatial coordinates ($s_1,s_3$) with spring stiffness (a) $\kappa_0 = 10^{-5}$ Nm$^{-1}$ and (b) $\kappa_0 = 10^{-2}$ Nm$^{-1}$. The radius of the two identically colliding flocs is $R = 1 \mu$m. The elastic ligands (or bonds with lower spring stiffness) have a larger contact area ($s_1, s_2$) with a non-zero collision impact.}\label{fig:g}
\end{figure}

In the absence of fluid flow, the bond ligand density, $g(D)$, is symmetric about the mean rest length of the binders, {\it l}$_0$ (Eqn.(\ref{eq:g_symm})). The adhesion mechanism is more efficient for elastic binders (i.e., springs with lower stiffness, $\kappa_0$), since these binders have a non-zero attachment over a larger contact area (e.g., compare the non-zero region in Fig. \ref{fig:g}a vs. Fig. \ref{fig:g}b). 



%
%
%

%
\noindent {\it (b) Floc-size distribution} -- To solve the complete population model (Eqn.(\ref{eq:Smol_final})) using the adhesion kernal described earlier, we employed the discretization scheme developed by Banks et. al. \cite{BanksKappel1989, AcklehFitzpatrick1997} and adopted by Doumic \cite{Prigent2012}. The parameters used in the simulations are listed in Table~\ref{tab:parameter}. 
\begin{table}[htbp]
\centering
\begin{tabular}{|c|c|c|c|c|}
\hline
Parameter & Value & Units & Source\\
\hline
$\kappa_0$ & (0.01-10) $\times 10^{-3}$ & N m$^{-1}$ & \cite{Mani2012}\\
\hline
${\it l}_0$ & $10^{-8}$ & m & \cite{Mani2012} \\
\hline
$\zeta$ &  (0.01-2.5) & N m$^{-1}$ s & -- \\
\hline
A$_{\text{Tot}}$ & 10$^9$ & m$^{-2}$ & \cite{Mani2012} \\
\hline
$\nicefrac{K^*_{\text{on}}}{K^*_{\text{off}}} $ & 10$^{-12}$ & -- & \cite{Mani2012} \\
\hline
$\gamma_A$ & 2.7 $\times 10^{-15}$ & fL$^{-2}$ & \cite{BortzEtal2008bmb} \\
\hline
\end{tabular}
\caption{Parameters common to all simulations.}\label{tab:parameter}
\end{table}
The convergence of the scheme was tested using the test functions in \cite{BortzEtal2008bmb}. A linear relationship between the L$^\infty$-error and the mesh-size, $\delta x$ was found using this first order approximation scheme.
%
%
%
%
%
%
The initial number density is chosen as $b_0(x) = 7.47 \times 10^{-4} e^{-0.00676x}$, where the coefficients are fit to the experimental data from the Younger Lab \cite{BortzEtal2008bmb}. The solutions in Fig.\ref{fig:b} are shown at time T=100 minutes. We chose 1 femtoliters (fL) as a lower bound $\underline{x}$ in our simulations. Our aggregation model allows the upper bound, $\overline{x}$, of the domain to go unrestrained (i.e., $\overline{x} \rightarrow \infty$), but the results are presented inside the window $1 \le x \le 1000$ fL.
%
%
%

%
Fig. \ref{fig:b} highlights the floc population at different surface parameter, $\kappa_0$, and fluid parameter, $\kappa$. These studies suggest that stiff binders lead to fewer large aggregates (i.e., b(x, T, $\kappa_0=10^{-2}$) $<$ b(x, T, $\kappa_0=10^{-3}$) $<$ b(x, T, $\kappa_0=10^{-5}$), for x $\ge 600$). This is not surprising since aggregation is influenced by the collision factor (see Eq. \ref{eq:KA}, \ref{eq:TotalForce}). A higher value of $g(D)$ suggests that two flocs close to each other are more likely to coalesce leading to bigger flocs. However, at a separation distance, $D=11$nm used in our numerical simulations, this factor is insignificant for stiff binders (e.g. compare the values of $g(D)$ in Fig. \ref{fig:g}a vs. Fig. \ref{fig:g}b) and does favor formation of large aggregates. Surface-adhesion is comparatively stronger in highly ionic fluids, represented by a shorter Debye length, $\kappa$. A short screening length implies a smaller separation distance between the interacting surfaces, and hence a strong adhesion (Fig. \ref{fig:b}b). Similarly, we have found that adhesion is favorable among flocs of smaller sizes (i.e., smaller radius of the coalescing spheres). This is effect is due to a smaller sphere-sphere potential energy barrier. 
\vskip 10pt
%
%


%
%
%
\noindent {\it Conclusion} -- We have presented a multi-scale model for the aggregation dynamics of rigid, charged, spherical, micron sized flocs. The equilibrium binding kinetics of the flocs is incorporated via the {\it collision factor}, a term popularly associated with the floc adhering efficiency, in the colloid literature \cite{Somasundaran}. Predictions about the floc aggregate size, at various fluid and surface potential parameters, are made using numerical simulations. Preliminary investigation in quiescent flow conditions highlight that the adhesion mechanism is favored if the binding ligands of the flocs are elastic, or the surrounding fluid is highly ionized, or the size of the aggregating flocs are sufficiently small. The effects of surface deformation which modifies the adhesion area and hence the aggregation kernal \cite{Jadhav2007}, spatial inhomogeneities of the material parameters, the non-equilibrium effects, stochasticity and the discrete number of bonds \cite{Zhu2000}, will certainly reveal interesting deviations in the macroscopic population dynamics. Hence these effects deserve deeper investigation in the future, especially on the experimental front.
\vskip 10pt

%

\noindent {\bf Acknowledgment:} This work supported in part by the National Science Foundation under grant NSF-1225878 and the National Institutes of Health under grant NIH-1R01GM081702-01A2.
\vskip -10pt

\bibliographystyle{apsrev}
\bibliography{/home/dmbortz/Desktop/pre/mathbioCUpre}

\begin{thebibliography}{35}
\expandafter\ifx\csname natexlab\endcsname\relax\def\natexlab#1{#1}\fi
\expandafter\ifx\csname bibnamefont\endcsname\relax
  \def\bibnamefont#1{#1}\fi
\expandafter\ifx\csname bibfnamefont\endcsname\relax
  \def\bibfnamefont#1{#1}\fi
\expandafter\ifx\csname citenamefont\endcsname\relax
  \def\citenamefont#1{#1}\fi
\expandafter\ifx\csname url\endcsname\relax
  \def\url#1{\texttt{#1}}\fi
\expandafter\ifx\csname urlprefix\endcsname\relax\def\urlprefix{URL }\fi
\providecommand{\bibinfo}[2]{#2}
\providecommand{\eprint}[2][]{\url{#2}}

\bibitem[{\citenamefont{Zhu}(2000)}]{Zhu2000}
\bibinfo{author}{\bibfnamefont{C.}~\bibnamefont{Zhu}},
  \bibinfo{journal}{Journal of Biomechanics} \textbf{\bibinfo{volume}{33}},
  \bibinfo{pages}{23} (\bibinfo{year}{2000}).

\bibitem[{\citenamefont{Lei et~al.}(1999)\citenamefont{Lei, Lawrence, and
  Dong}}]{Lei1999}
\bibinfo{author}{\bibfnamefont{X.}~\bibnamefont{Lei}},
  \bibinfo{author}{\bibfnamefont{M.~B.} \bibnamefont{Lawrence}},
  \bibnamefont{and} \bibinfo{author}{\bibfnamefont{C.}~\bibnamefont{Dong}},
  \bibinfo{journal}{J. Biomech. Eng.} \textbf{\bibinfo{volume}{121}},
  \bibinfo{pages}{636} (\bibinfo{year}{1999}).

\bibitem[{\citenamefont{Moore and Kuhl}(2006)}]{Moore2006}
\bibinfo{author}{\bibfnamefont{N.~W.} \bibnamefont{Moore}} \bibnamefont{and}
  \bibinfo{author}{\bibfnamefont{T.~L.} \bibnamefont{Kuhl}},
  \bibinfo{journal}{Biophys. J.} \textbf{\bibinfo{volume}{91}},
  \bibinfo{pages}{1675} (\bibinfo{year}{2006}).

\bibitem[{\citenamefont{Somasundaran et~al.}(2005)\citenamefont{Somasundaran,
  Runkanan, Kapur, Stechemesser, and Dobi\'{a}\v{s}}}]{Somasundaran}
\bibinfo{author}{\bibfnamefont{P.}~\bibnamefont{Somasundaran}},
  \bibinfo{author}{\bibfnamefont{V.}~\bibnamefont{Runkanan}},
  \bibinfo{author}{\bibfnamefont{P.}~\bibnamefont{Kapur}},
  \bibinfo{author}{\bibfnamefont{H.}~\bibnamefont{Stechemesser}},
  \bibnamefont{and}
  \bibinfo{author}{\bibfnamefont{B.}~\bibnamefont{Dobi\'{a}\v{s}}}, in
  \emph{\bibinfo{booktitle}{Coagulation and Flocculation}}
  (\bibinfo{publisher}{Taylor \& Francis}, \bibinfo{year}{2005}), vol.
  \bibinfo{volume}{126}, chap.~\bibinfo{chapter}{11}, pp.
  \bibinfo{pages}{767--803}, \bibinfo{edition}{2nd} ed.

\bibitem[{\citenamefont{Maximova and Dahl}(2006)}]{Maximova2006}
\bibinfo{author}{\bibfnamefont{N.}~\bibnamefont{Maximova}} \bibnamefont{and}
  \bibinfo{author}{\bibfnamefont{O.}~\bibnamefont{Dahl}},
  \bibinfo{journal}{Current Opinion in Colloid and Interface Science}
  \textbf{\bibinfo{volume}{11}}, \bibinfo{pages}{246} (\bibinfo{year}{2006}).

\bibitem[{\citenamefont{Vogelaar et~al.}(2005)\citenamefont{Vogelaar, {De
  Keizer}, Spijker, and Lettinga}}]{Vogelaar2005}
\bibinfo{author}{\bibfnamefont{J.~C.~T.} \bibnamefont{Vogelaar}},
  \bibinfo{author}{\bibfnamefont{A.}~\bibnamefont{{De Keizer}}},
  \bibinfo{author}{\bibfnamefont{S.}~\bibnamefont{Spijker}}, \bibnamefont{and}
  \bibinfo{author}{\bibfnamefont{G.}~\bibnamefont{Lettinga}},
  \bibinfo{journal}{Water research} \textbf{\bibinfo{volume}{39}},
  \bibinfo{pages}{37} (\bibinfo{year}{2005}).

\bibitem[{\citenamefont{Mohapatra et~al.}(2010)\citenamefont{Mohapatra, Brar,
  Tyagi, and Surampalli}}]{Mohapatra2010}
\bibinfo{author}{\bibfnamefont{D.}~\bibnamefont{Mohapatra}},
  \bibinfo{author}{\bibfnamefont{S.}~\bibnamefont{Brar}},
  \bibinfo{author}{\bibfnamefont{R.}~\bibnamefont{Tyagi}}, \bibnamefont{and}
  \bibinfo{author}{\bibfnamefont{R.}~\bibnamefont{Surampalli}},
  \bibinfo{journal}{Chemical Engineering Journal}
  \textbf{\bibinfo{volume}{163}}, \bibinfo{pages}{273} (\bibinfo{year}{2010}).

\bibitem[{\citenamefont{Dembo et~al.}(1988)\citenamefont{Dembo, Torney, Saxman,
  and Hammer}}]{Dembo1988}
\bibinfo{author}{\bibfnamefont{M.}~\bibnamefont{Dembo}},
  \bibinfo{author}{\bibfnamefont{D.~C.} \bibnamefont{Torney}},
  \bibinfo{author}{\bibfnamefont{K.}~\bibnamefont{Saxman}}, \bibnamefont{and}
  \bibinfo{author}{\bibfnamefont{D.}~\bibnamefont{Hammer}},
  \bibinfo{journal}{Proc. Royal Soc. B} \textbf{\bibinfo{volume}{234}},
  \bibinfo{pages}{55} (\bibinfo{year}{1988}).

\bibitem[{\citenamefont{Bell}(1978)}]{Bell1978}
\bibinfo{author}{\bibfnamefont{G.}~\bibnamefont{Bell}},
  \bibinfo{journal}{Science} \textbf{\bibinfo{volume}{200}},
  \bibinfo{pages}{618} (\bibinfo{year}{1978}).

\bibitem[{\citenamefont{Reboux et~al.}(2008)\citenamefont{Reboux, Richardson,
  and Jensen}}]{Reboux2008}
\bibinfo{author}{\bibfnamefont{S.}~\bibnamefont{Reboux}},
  \bibinfo{author}{\bibfnamefont{G.}~\bibnamefont{Richardson}},
  \bibnamefont{and} \bibinfo{author}{\bibfnamefont{O.}~\bibnamefont{Jensen}},
  \bibinfo{journal}{Proc. Royal Soc. A} \textbf{\bibinfo{volume}{464}},
  \bibinfo{pages}{447} (\bibinfo{year}{2008}).

\bibitem[{\citenamefont{Evans}(1985)}]{Evans1985}
\bibinfo{author}{\bibfnamefont{E.~A.} \bibnamefont{Evans}},
  \bibinfo{journal}{Biophys. J.} \textbf{\bibinfo{volume}{48}},
  \bibinfo{pages}{175} (\bibinfo{year}{1985}).

\bibitem[{\citenamefont{Goldman et~al.}(1967)\citenamefont{Goldman, Cox, and
  Brenner}}]{GoldmanCoxBrenner1967CES}
\bibinfo{author}{\bibfnamefont{A.}~\bibnamefont{Goldman}},
  \bibinfo{author}{\bibfnamefont{R.}~\bibnamefont{Cox}}, \bibnamefont{and}
  \bibinfo{author}{\bibfnamefont{H.}~\bibnamefont{Brenner}},
  \bibinfo{journal}{Chemical Engineering Science}
  \textbf{\bibinfo{volume}{22}}, \bibinfo{pages}{637} (\bibinfo{year}{1967}).

\bibitem[{\citenamefont{King and Hammer}(2001)}]{King2001}
\bibinfo{author}{\bibfnamefont{M.~R.} \bibnamefont{King}} \bibnamefont{and}
  \bibinfo{author}{\bibfnamefont{D.~A.} \bibnamefont{Hammer}},
  \bibinfo{journal}{Biophys. J.} \textbf{\bibinfo{volume}{81}},
  \bibinfo{pages}{799} (\bibinfo{year}{2001}).

\bibitem[{\citenamefont{Korn and Schwarz}(2006)}]{Korn2006}
\bibinfo{author}{\bibfnamefont{C.}~\bibnamefont{Korn}} \bibnamefont{and}
  \bibinfo{author}{\bibfnamefont{U.~S.} \bibnamefont{Schwarz}},
  \bibinfo{journal}{Physical Rev. Letters} \textbf{\bibinfo{volume}{97}},
  \bibinfo{pages}{138103} (\bibinfo{year}{2006}).

\bibitem[{\citenamefont{Mani et~al.}(2012)\citenamefont{Mani, Gopinath, and
  Mahadevan}}]{Mani2012}
\bibinfo{author}{\bibfnamefont{M.}~\bibnamefont{Mani}},
  \bibinfo{author}{\bibfnamefont{A.}~\bibnamefont{Gopinath}}, \bibnamefont{and}
  \bibinfo{author}{\bibfnamefont{L.}~\bibnamefont{Mahadevan}},
  \bibinfo{journal}{Physical Review Letters} \textbf{\bibinfo{volume}{108}}
  (\bibinfo{year}{2012}).

\bibitem[{\citenamefont{Bihr et~al.}(2012)\citenamefont{Bihr, Seifert, and
  Smith}}]{Bihr2012}
\bibinfo{author}{\bibfnamefont{T.}~\bibnamefont{Bihr}},
  \bibinfo{author}{\bibfnamefont{U.}~\bibnamefont{Seifert}}, \bibnamefont{and}
  \bibinfo{author}{\bibfnamefont{A.~S.} \bibnamefont{Smith}},
  \bibinfo{journal}{Physical Rev. Letters} \textbf{\bibinfo{volume}{109}},
  \bibinfo{pages}{258101} (\bibinfo{year}{2012}).

\bibitem[{\citenamefont{Cogan and Keener}(2004)}]{Cogan2004}
\bibinfo{author}{\bibfnamefont{N.~G.} \bibnamefont{Cogan}} \bibnamefont{and}
  \bibinfo{author}{\bibfnamefont{J.~P.} \bibnamefont{Keener}},
  \bibinfo{journal}{Mathematical Medicine and Biology}
  \textbf{\bibinfo{volume}{21}}, \bibinfo{pages}{147} (\bibinfo{year}{2004}).

\bibitem[{\citenamefont{T.~Zhang and Wang}(2008)}]{Zhang2008}
\bibinfo{author}{\bibfnamefont{N.~C.} \bibnamefont{T.~Zhang}} \bibnamefont{and}
  \bibinfo{author}{\bibfnamefont{Q.}~\bibnamefont{Wang}},
  \bibinfo{journal}{Comm. Comp. Phys.} \textbf{\bibinfo{volume}{4}},
  \bibinfo{pages}{72} (\bibinfo{year}{2008}).

\bibitem[{\citenamefont{S.~Corezzi and Sciortino}(2012)}]{Corezzi2012}
\bibinfo{author}{\bibfnamefont{D.~F.} \bibnamefont{S.~Corezzi}}
  \bibnamefont{and}
  \bibinfo{author}{\bibfnamefont{F.}~\bibnamefont{Sciortino}},
  \bibinfo{journal}{Soft Matter} \textbf{\bibinfo{volume}{8}},
  \bibinfo{pages}{11207} (\bibinfo{year}{2012}).

\bibitem[{\citenamefont{Sokurenko et~al.}(1997)\citenamefont{Sokurenko,
  Chesnokova, Doyle, and Hasty}}]{Sokurenko1997}
\bibinfo{author}{\bibfnamefont{E.~V.} \bibnamefont{Sokurenko}},
  \bibinfo{author}{\bibfnamefont{V.}~\bibnamefont{Chesnokova}},
  \bibinfo{author}{\bibfnamefont{R.~J.} \bibnamefont{Doyle}}, \bibnamefont{and}
  \bibinfo{author}{\bibfnamefont{D.~L.} \bibnamefont{Hasty}},
  \bibinfo{journal}{The Journal of biological chemistry}
  \textbf{\bibinfo{volume}{272}}, \bibinfo{pages}{17880}
  (\bibinfo{year}{1997}).

\bibitem[{\citenamefont{Sokurenko et~al.}(1998)\citenamefont{Sokurenko,
  Chesnokova, Dykhuizen, Ofek, Wu, Krogfelt, Struve, Schembri, and
  Hasty}}]{Sokurenko1998}
\bibinfo{author}{\bibfnamefont{E.~V.} \bibnamefont{Sokurenko}},
  \bibinfo{author}{\bibfnamefont{V.}~\bibnamefont{Chesnokova}},
  \bibinfo{author}{\bibfnamefont{D.~E.} \bibnamefont{Dykhuizen}},
  \bibinfo{author}{\bibfnamefont{I.}~\bibnamefont{Ofek}},
  \bibinfo{author}{\bibfnamefont{X.~R.} \bibnamefont{Wu}},
  \bibinfo{author}{\bibfnamefont{K.~A.} \bibnamefont{Krogfelt}},
  \bibinfo{author}{\bibfnamefont{C.}~\bibnamefont{Struve}},
  \bibinfo{author}{\bibfnamefont{M.~A.} \bibnamefont{Schembri}},
  \bibnamefont{and} \bibinfo{author}{\bibfnamefont{D.~L.} \bibnamefont{Hasty}},
  \bibinfo{journal}{Proceedings of the National Academy of Sciences}
  \textbf{\bibinfo{volume}{95}}, \bibinfo{pages}{8922} (\bibinfo{year}{1998}).

\bibitem[{\citenamefont{Cooper}(1997)}]{Cooper1997}
\bibinfo{author}{\bibfnamefont{S.}~\bibnamefont{Cooper}}, \bibinfo{journal}{J.
  Bacteriology} \textbf{\bibinfo{volume}{179}}, \bibinfo{pages}{5582}
  (\bibinfo{year}{1997}).

\bibitem[{\citenamefont{Tuson et~al.}(2012)\citenamefont{Tuson, Auer, Renner,
  Hasebe, Tropini, Salick, Crone, Gopinathan, Huang, and Weibel}}]{Tuson2012}
\bibinfo{author}{\bibfnamefont{H.~H.} \bibnamefont{Tuson}},
  \bibinfo{author}{\bibfnamefont{G.~K.} \bibnamefont{Auer}},
  \bibinfo{author}{\bibfnamefont{L.~D.} \bibnamefont{Renner}},
  \bibinfo{author}{\bibfnamefont{M.}~\bibnamefont{Hasebe}},
  \bibinfo{author}{\bibfnamefont{C.}~\bibnamefont{Tropini}},
  \bibinfo{author}{\bibfnamefont{M.}~\bibnamefont{Salick}},
  \bibinfo{author}{\bibfnamefont{W.~C.} \bibnamefont{Crone}},
  \bibinfo{author}{\bibfnamefont{A.}~\bibnamefont{Gopinathan}},
  \bibinfo{author}{\bibfnamefont{K.~C.} \bibnamefont{Huang}}, \bibnamefont{and}
  \bibinfo{author}{\bibfnamefont{D.~B.} \bibnamefont{Weibel}},
  \bibinfo{journal}{Molecular Microbiology} \textbf{\bibinfo{volume}{84}},
  \bibinfo{pages}{874} (\bibinfo{year}{2012}).

\bibitem[{\citenamefont{Marshall et~al.}(2003)\citenamefont{Marshall, Long,
  Piper, Yago, and McEver}}]{Marshall2003}
\bibinfo{author}{\bibfnamefont{B.}~\bibnamefont{Marshall}},
  \bibinfo{author}{\bibfnamefont{M.}~\bibnamefont{Long}},
  \bibinfo{author}{\bibfnamefont{J.}~\bibnamefont{Piper}},
  \bibinfo{author}{\bibfnamefont{T.}~\bibnamefont{Yago}}, \bibnamefont{and}
  \bibinfo{author}{\bibfnamefont{R.}~\bibnamefont{McEver}},
  \bibinfo{journal}{Nature} \textbf{\bibinfo{volume}{423}},
  \bibinfo{pages}{190} (\bibinfo{year}{2003}).

\bibitem[{\citenamefont{Thomas}(2008)}]{Thomas2008}
\bibinfo{author}{\bibfnamefont{W.}~\bibnamefont{Thomas}},
  \emph{\bibinfo{title}{Catch bonds in adhesion}}, vol.~\bibinfo{volume}{10}
  (\bibinfo{publisher}{Annual Reviews}, \bibinfo{year}{2008}).

\bibitem[{\citenamefont{Byrne et~al.}(2011)\citenamefont{Byrne, Dzul, Solomon,
  Younger, and Bortz}}]{Byrneetal2011PRE}
\bibinfo{author}{\bibfnamefont{E.~C.} \bibnamefont{Byrne}},
  \bibinfo{author}{\bibfnamefont{S.~P.} \bibnamefont{Dzul}},
  \bibinfo{author}{\bibfnamefont{M.~J.} \bibnamefont{Solomon}},
  \bibinfo{author}{\bibfnamefont{J.~G.} \bibnamefont{Younger}},
  \bibnamefont{and} \bibinfo{author}{\bibfnamefont{D.~M.} \bibnamefont{Bortz}},
  \bibinfo{journal}{Physical Review E} \textbf{\bibinfo{volume}{83}},
  \bibinfo{pages}{41911} (\bibinfo{year}{2011}).

\bibitem[{\citenamefont{Bortz et~al.}(2008)\citenamefont{Bortz, Jackson,
  Taylor, Thompson, and Younger}}]{BortzEtal2008bmb}
\bibinfo{author}{\bibfnamefont{D.~M.} \bibnamefont{Bortz}},
  \bibinfo{author}{\bibfnamefont{T.~L.} \bibnamefont{Jackson}},
  \bibinfo{author}{\bibfnamefont{K.~A.} \bibnamefont{Taylor}},
  \bibinfo{author}{\bibfnamefont{A.~P.} \bibnamefont{Thompson}},
  \bibnamefont{and} \bibinfo{author}{\bibfnamefont{J.~G.}
  \bibnamefont{Younger}}, \bibinfo{journal}{Bull. Math. Biology}
  \textbf{\bibinfo{volume}{70}}, \bibinfo{pages}{745} (\bibinfo{year}{2008}).

\bibitem[{\citenamefont{Tabatabaei and Ven}(2010)}]{Tabatabaei2010}
\bibinfo{author}{\bibfnamefont{S.}~\bibnamefont{Tabatabaei}} \bibnamefont{and}
  \bibinfo{author}{\bibfnamefont{T.~V.~D.} \bibnamefont{Ven}},
  \bibinfo{journal}{J. Fluid Mechanics} \textbf{\bibinfo{volume}{656}},
  \bibinfo{pages}{360} (\bibinfo{year}{2010}).

\bibitem[{\citenamefont{Abu-Lail and Camesano}(2003)}]{Abu-Lail2003}
\bibinfo{author}{\bibfnamefont{N.~I.} \bibnamefont{Abu-Lail}} \bibnamefont{and}
  \bibinfo{author}{\bibfnamefont{T.~A.} \bibnamefont{Camesano}},
  \bibinfo{journal}{Biomacromolecules} \textbf{\bibinfo{volume}{4}},
  \bibinfo{pages}{1000} (\bibinfo{year}{2003}).

\bibitem[{\citenamefont{Israelachvili}(2011)}]{Israelachvili2011}
\bibinfo{author}{\bibfnamefont{J.}~\bibnamefont{Israelachvili}},
  \emph{\bibinfo{title}{{Intermolecular and Surface Forces}}}
  (\bibinfo{publisher}{Academic Press}, \bibinfo{address}{Amsterdam},
  \bibinfo{year}{2011}), \bibinfo{edition}{3rd} ed.

\bibitem[{\citenamefont{Gregory}(2006)}]{Gregory2006}
\bibinfo{author}{\bibfnamefont{J.}~\bibnamefont{Gregory}},
  \emph{\bibinfo{title}{{Particles in Water}}} (\bibinfo{publisher}{CRC Press},
  \bibinfo{year}{2006}).

\bibitem[{\citenamefont{Banks and Kappel}(1989)}]{BanksKappel1989}
\bibinfo{author}{\bibfnamefont{H.~T.} \bibnamefont{Banks}} \bibnamefont{and}
  \bibinfo{author}{\bibfnamefont{F.}~\bibnamefont{Kappel}},
  \bibinfo{journal}{Semigroup Forum} \textbf{\bibinfo{volume}{38}},
  \bibinfo{pages}{141} (\bibinfo{year}{1989}).

\bibitem[{\citenamefont{Ackleh and Fitzpatrick}(1997)}]{AcklehFitzpatrick1997}
\bibinfo{author}{\bibfnamefont{A.~S.} \bibnamefont{Ackleh}} \bibnamefont{and}
  \bibinfo{author}{\bibfnamefont{B.~G.} \bibnamefont{Fitzpatrick}},
  \bibinfo{journal}{J. Math. Biology} \textbf{\bibinfo{volume}{35}},
  \bibinfo{pages}{480} (\bibinfo{year}{1997}).

\bibitem[{\citenamefont{Prigent et~al.}(2012)\citenamefont{Prigent, Ballesta,
  Charles, Lenuzza, Gabriel, Tine, Rezaei, and Doumic}}]{Prigent2012}
\bibinfo{author}{\bibfnamefont{S.}~\bibnamefont{Prigent}},
  \bibinfo{author}{\bibfnamefont{A.}~\bibnamefont{Ballesta}},
  \bibinfo{author}{\bibfnamefont{F.}~\bibnamefont{Charles}},
  \bibinfo{author}{\bibfnamefont{N.}~\bibnamefont{Lenuzza}},
  \bibinfo{author}{\bibfnamefont{P.}~\bibnamefont{Gabriel}},
  \bibinfo{author}{\bibfnamefont{L.~M.} \bibnamefont{Tine}},
  \bibinfo{author}{\bibfnamefont{H.}~\bibnamefont{Rezaei}}, \bibnamefont{and}
  \bibinfo{author}{\bibfnamefont{M.}~\bibnamefont{Doumic}},
  \bibinfo{journal}{PLoS ONE} \textbf{\bibinfo{volume}{7}},
  \bibinfo{pages}{e43273} (\bibinfo{year}{2012}).

\bibitem[{\citenamefont{Jadhav et~al.}(2007)\citenamefont{Jadhav, Eggleton, and
  Konstantopoulos}}]{Jadhav2007}
\bibinfo{author}{\bibfnamefont{S.}~\bibnamefont{Jadhav}},
  \bibinfo{author}{\bibfnamefont{C.~D.} \bibnamefont{Eggleton}},
  \bibnamefont{and}
  \bibinfo{author}{\bibfnamefont{K.}~\bibnamefont{Konstantopoulos}},
  \bibinfo{journal}{Current pharmaceutical design}
  \textbf{\bibinfo{volume}{13}}, \bibinfo{pages}{1511} (\bibinfo{year}{2007}).

\end{thebibliography}

\end{document}